\documentclass[10pt]{article} 
\usepackage{ragged2e}
\usepackage[utf8]{inputenc} 
\usepackage[pdftex]{graphicx}

\usepackage{geometry} 
\usepackage{graphicx} 
\usepackage{amsmath}
\usepackage{mathtools}
\usepackage{upgreek}

\usepackage{booktabs} 
\usepackage{array} 
\usepackage{paralist} 
\usepackage{verbatim} 
\usepackage{subfig} 

\usepackage{fancyhdr} 
\usepackage{sectsty}
\allsectionsfont{\sffamily\mdseries\upshape} 

\usepackage[nottoc,notlof,notlot]{tocbibind} 
\usepackage[titles,subfigure]{tocloft} 

\title{On the Use of Dimension Reduction or Signal Separation Methods for Nitrogen River Pollution Source Identification}
\author{Hatipoğlu, Güray}
\date{April 27, 2022} 

\begin{document}
\maketitle

\centering{\section*{Abstract}}
\justifying
Identification of the current and expected future pollution sources to rivers is crucial for sound environmental management. For this purpose numerous approaches were proposed that can be clustered under physical based models, stable isotope analysis and mixing methods, mass balance methods, time series analysis, land cover analysis, and spatial statistics. Another extremely common method is Principal Component Analysis, as well as its modifications, such as Absolute Principal Component Score. they have been applied to the source identification problems for nitrogen entry to rivers. This manuscript is checking whether PCA can really be a powerful method to uncover nitrogen pollution sources considering its theoretical background and assumptions. Moreover, slightly similar techniques, Independent Component Analysis and Factor Analysis will also be considered.

\setlength{\columnsep}{3em}
\setlength{\columnseprule}{1pt}

\centering\section{Introduction}
\justifying
The excess nitrogen load into aquatic environment result in unwanted ecological conditions, as well as nitrogenous greenhouse gas emission and valuable reactive nitrogen source. Among the mediums, rivers are very significant. They usually do not have very high capacity to accumulate nitrogen to excessive concentrations by themselves, yet they can carry it to other environments, coastal or inland, to their predicament (Galloway, 2005). Governing this issue requires working knowledge over how nitrogen behaves in the environment and what are the loads of nitrogen to a specific region. 

Nitrogen can reside in all mediums. A summary of river relevant inorganic nitrogen species is given in the following table. 
\newpage
\begin{center}
\centering\captionof{table}{Relevant inorganic nitrogen species for river environments}
\begin{tabular}{ c c c }
\hline
Name & Formula & Molecular Mass (g/mol) \\ 
Ammonium & NH\textsubscript{4}\textsuperscript{+} & 17 \\  
Dinitrogen & N\textsubscript{2} & 28 \\
Nitrate & NO\textsubscript{3}\textsuperscript{-} & 62 \\
Nitrite & NO\textsubscript{2}\textsuperscript{-} & 46 \\
Nitrogen Dioxide & NO\textsubscript{2} & 78 \\
Nitrogen Monoxide & NO & 62 \\
Nitrous Oxide & N\textsubscript{2}O & 44  \\
\hline
\label{Table. 1} 
\end{tabular}
\end{center}
Other relevant environmental measurements for nitrogen are Total Nitrogen (N (from ammonia, N-NH4) + N-NO3 + N-NO2 (from nitrite) + organic N) and Total Kjeldahl Nitrogen (Organic N + N-NH4). They will be referred at the later parts of this manuscript again.\\

Atmosphere mostly has inert dinitrogen, where biological or technological fixation can make it available to plants and others, in addition to the lightning bolts. Atmosphere also transports and deposits land-generated combustion byproduct nitrous oxides, evaporated ammonia, denitrified nitrous oxide and dinitrogen as well as their counterparts from aquatic environments. Very high majority of the plants can only utilize nitrate and partially nitrite and ammonium, while only a fraction of photosynthetic cyanobacteria and several other bacteria type can fix dinitrogen directly. Consequently, ammonium and nitrate concentrations are monitored to estimate state of the aquatic ecosystem and possible eutrophication or harmful algal bloom conditions. More complex transformation of these chemical species can be simulated in biogeochemical models to understand nitrogen fate and transport better. \\

\begin{figure}
\includegraphics{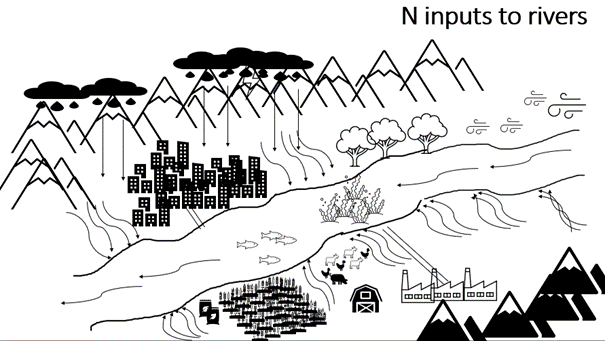}
\caption{Main ways of entry to rivers by nitrogen species. From left top to right top; atmospheric wet deposition, sewage discharge, urban runoff, atmospheric dry deposition, upstream river flow. From bottom right to bottom left; soil decomposition, industrial wastewater discharge, agricultural runoff (manure), agricultural runoff (fertilizer)}
\label{Fig. 1} 
\end{figure}

\justifying
The part above introduced the basic movements of nitrogenous species through the environment. In the following section, a close-up view to riverine dynamics of nitrate will be examined to permit the selection of a better algorithm to make sense of collected \textit{in situ} data.

\centering\section{What is Happening While Nitrogen is Polluting Rivers?}
\justifying
The nitrogen dynamics in rivers is simply too complex to satisfactorily include here as a whole. However, the critical parts will be tried to be included to be able to select the necessary statistical tool for later analyses of source apportionment. For a full investigation of global nitrogen cycle, the readers are kindly referred to Ussiri and Lal (2013).

One complication of nitrogen dynamics in river is that many other seemingly unrelated components can directly or indirectly impact its fate. Phytoplankton and cyanobacteria both consumes and generates nitrogen species while they are affected by many other environmental conditions, including but not limited to salinity, temperature, phosphorus. Their numbers, species and spatial distributions vary with time even without considering their predators, such as zooplanktons, which further complicates the situation. Changes in river dynamics, flows, surface run-off and groundwater inflows to rivers also impacts both concentration and amount of nitrogen species. As a result, many water quality variables and physical characteristics are interdependent, hence if not impossible, not causally related variable set generation is very difficult, let alone simple covariance/correlation. Henceforth, the first remark is, \textbf{the approaches assuming that water quality and physical characteristics are independent from each other are not suitable for nitrogen source apportionment}. 
Another issue is spatial and temporal correlation between nitrate measurement values of samples collected throughout the study area in rivers. Many statistical analyses and machine learning routines work much better with larger data volumes. However, the samples fed to the algorithms are in some degree not independent from each other. In line with intuition, upstream nearby sampling point is similar to the downstream ones. From the temporal perspective, there are daily, seasonally, and yearly cycles in many nitrogen emitting processes. There are several ways to check and act according to these intersample correlations. For instance, Moran's I value related calculations have been used to determine minimum distance between river sampling network that will result in \textit{spatially uncorrelated} locations. However, the fact that many point and non-point pollution sources enter the river from different points, and consequent differences in different set of nitrogen sources in different locations should be kept in mind. For temporal analyses, there are various decomposition techniques to get rid of temporal autocorrelations. Yet, one critical point is the complexity of source apportionment will be higher if the proportions of those sources on their nitrogen contribution to the rivers changes significantly with time. Another possible way is taking a "temporal snapshot" while considering spatial autocorrelation as well, and make separate analyses, with obvious trade-offs from data volume. In short, \textbf{the individual measurement points, or simply sample measurements' independency from each other should be thoroughly checked prior to the analyses, especially the analyses assuming that samples are independent.}\\
Yet another point is related to the sources and sinks. Moreover, there are several factors that affect different sources in varying degrees. The following examples will try to clarify these statements: \\
Urban runoff and sewage discharge are both related to the size of city and consequently to population casually. Agricultural runoff from manure and fertilizer can be correlated since manure is also being used as a type of organic fertilizer. The nitrogen from the atmospheric deposition is, in addition to the combustion and lightning bolts, sourced by nearby agricultural nitrogen emissions as well. \\
For factors affecting many of these sources, the first one is changes in precipitation. Surface runoff from urban or agricultural environments are either creathed or enhanced by precipitation. Wastewater discharges can be diluted by precipitation, or after exceeding the capacity of treatment facilities it might result in untreated wastewater discharge. Wet deposition is directly coming with precipitation.
The second factor is temperature. If it results in a decline in relative humidity, crops will draw more water, hence fertilizer loss might be lower. On the other hand, overall increase in aerobic bacterial activity will decompose soil organic matter faster, hence more nitrogen loss from this source. Additionally, increase in water temperature will also increase \textit{denitrification}, which will reduce nitrate level overall in water.
Last factor is related to environmental management in a municipal or regional level. Urban runoffs, sewerage, wastewater treatment plants and other relevant infrastructures may or may not be robust enough to constrain the adverse impacts of anthropogenic pollution sources. Henceforth, \textbf{pollution sources (as well as sinks) can be directly or indirectly related to each other, they are not independent}. \\
Nevertheless, a number of techniques has been developed to date with non-trivial successes that can still be able to predict the nitrogen pollution in the midst of all this complexity. The following three sections will explain three relatively similar methods, Principal Component Analysis, Independent Component Analysis, and Factor Analysis one by one with now their assumptions. In this way, the suitability conditions of these methods for the case of nitrogen pollution source apportionment will be discussed.

\centering\section{What Does Principal Component Analysis (PCA) Do?}
\justifying
PCA is generally known as a \textit{dimensional reduction} method or a \textit{way to rotate axises to maximize variance}. To comprehend the theory behind the PCA and its advantages/shortcomings, readers are referred to the tutorial of Shlens (2014) and book of Everitt and Hothorn (2011). Several points of PCA are considered below: \\
\textit{Covariance matrix construction}\\
Assume that we have a dataset with N samples and M variables. Firstly, mean of each variable will be subtracted from its corresponding sample value for every sample. After this, how the variables \textit{covary} with each other will be examined via covariance matrix.
\\
\\
\[
\begin{bmatrix}
\sigma_{1}\sigma_{1}&\sigma_{1}\sigma_{2}&\sigma_{1}\sigma_{3} & \sigma_{1}\sigma_{4}\\
\sigma_{2}\sigma_{1}&\sigma_{2}\sigma_{2}&\sigma_{2}\sigma_{3} & \sigma_{2}\sigma_{4}\\
\sigma_{3}\sigma_{1}&\sigma_{3}\sigma_{2}&\sigma_{3}\sigma_{3} & \sigma_{3}\sigma_{4}\\
\sigma_{4}\sigma_{1}&\sigma_{4}\sigma_{2}&\sigma_{4}\sigma_{3} & \sigma_{4}\sigma_{4}
\end{bmatrix}
\]
\\
\\
\justifying
\\
These are standard deviations of i and y variables, and their multiplication results in covariance of i and y variables. If two variables vary together the multiplication of their standard deviations will be nearer to 1, and nearer to 0 in the opposite case. As this is a very powerful way to check the overall situation between variables and that many statistical methods have this in their processes, R Statistical software, MATLAB and Python languages have a single function to create it. After this step, creation the correlation matrix of the same dataset is easy, permitting to check the traces of multicollinearity among variables. To the covariance matrix (or correlation matrix) a matrix decomposition method, eigenvalue decomposition can be applied. The entire dataset will be rotated from the original variables to a new space with \textit{principal components} ordered from maximum explanation of variance in data to the lowest one, which is in general noise in the data. Each principal component is a linear combinations of variables with different weights\\
The actual pollution source apportionment part comes now. Numerous research has been done via directly interpreting these different principal components as \textit{sources that vary each variable in a certain degree}, with high weights correspond to higher dominance by a certain source. The main intuition behind this approach is that if a group of variables are strongly and significantly covary according to a single pirncipal component, then they are ought to be impacted by a certain source that simultaneously introduce them or directly/indirectly affect them together. Everitt and Hothorn found this as an overinterpretation, as PCA is not and does not do what is directly expected from a source apportionment algorithm. Albeit many researchers  reported quite high variance explanation and well-sounding principal component to source elaborations, PCA is not itself appropriate for non-linear cases, and sources and changes in nitrogen dynamics are frequently non-linear in nature, with mentioned spatio-temporal autocorrelations. To encounter these, \\ 1-) many other modified versions of PCA (kernel PCA etc.) might be used. An introduction to a comprehensive discussion over PCA itself and comments on this discussion can be found in Joliffe's book (2002), or \\ 2-) variables might be aggregated (see Total Nitrogen above) or discarded, or \\ 3-) other methods that are insensitive against mentioned limitations might be selected, or \\ 4-) samples might be reduced to get rid of some autocorrelations, or decomposed, or \\ 5-) in case temporal data is used, for instance on a single station, rather than the use of actual nitrate concentration values, difference of nitrate concentration between two consecutive time-steps might be used. \\ These considerations (except 3) will be further discussed in the following application seciton, after considering ICA and FA approaches. \\

\centering\section{Can Independent Component Analysis (ICA) be More Appropriate?}
\justifying
Independent Component Analysis, as its name implies, looks for independent components and it is directly being used for source apportionment, \textit{blind source separation}. Readers are kindly referred to Shlens (2014b) tutorial over how it works, theoretical basis and comparisons with PCA.\\
The base under the ICA is Singular Value Decomposition.

\begin{equation}
A=UAV^T
\end{equation}
\\

First half of the ICA methdology is as if doing PCA and normalizing the variances to have dimensions in standard units. This removes linear dependencies. Later, statistical independence will be ensured via putting a constraint on data from information theory. Here, the aim is making the statistical dependence near-zero: \\

\begin{equation}
I(y)=\int{P(y)log_{2}\frac{P(y)} {\prod_{i=i}^{} P(y_{i})}  dy}
\end{equation}

This very strict criterion also result in removal of higher-order dependencies in the dataset compared to the PCA. The critical point is because of the "requirement" of non-gaussian distributions in the dataset, there are local minima in the independent component searching second part. Infact, ICA produces similar results to PCA when the distribution of data is Gaussian. The readers are referred to the book of \textit{Independent Component Analysis} by Hyvarinen, Karhunen and Oja (2001)'s page 161 for the caveat of this non-Gaussian criterion.

\centering\section{Is This Topic More Suited to be Studied by Exploratory Factor Analysis?}
\justifying
FA simply analyzes the factors, i.e. \textit{latent variables}, over the dataset. The latent variable is hard to directly measure, while its effects can be measured and later related to latent variable quantitatively and linearly. For the theoretical basis and examples on its use, the \textit{Exploratory Factor Analysis} and \textit{Confirmatory Factor Analysis and Structural Equation Models} chapters of Everitt and Hothorn (2011), and Adachi (2016) are recommended. 
Several caveats in the factor analysis as follow: Factor analysis can be made via "principal factor analysis", similar to PCA but focusing on \textbf{common} factors and tries the find the variance that is related to other variables as well. Another way to achieve it is maximum likelihood method. \\
Overall, what FA assumes in the actual situartion is below: Let A, B, C, D, E are variables, and Factor1 and Factor2 are \textit{latent variables}, or factors, "c"s are intercepts of the equations and "e"s are error terms.
\\
\begin{align*} 
A = \textit{a}_{11}   \times   Factor_1 + \textit{a}_{12}   \times   Factor_2 + c_{1} + e_{1} \\
B = \textit{a}_{21}   \times   Factor_1 + \textit{a}_{22}   \times   Factor_2 + c_{2} + e_{2} \\
C = \textit{a}_{31}   \times   Factor_1 + \textit{a}_{32}   \times   Factor_2 + c_{3} + e_{3} \\
D = \textit{a}_{41}   \times   Factor_1 + \textit{a}_{42}   \times   Factor_2 + c_{4} + e_{4} \\
E = \textit{a}_{51}   \times   Factor_1 + \textit{a}_{52}   \times   Factor_2 + c_{5} + e_{5} \\
\end{align*}
\\
\justifying
The error term \textit{e} has its own variance for each variable, while the Factors are common with all variables with only different weights. This separation is quite important as the FA look for \textit{common} factors/variances and not the unique variance of each variable

Determining the number of factors to be retrieved is \textit{a priori}, and the same dataset will generate different factors and variable loadings with changing number of factors. Slightly similar to the k-means clustering, one way is starting form a low number of factor and keep increasing the number until the minimum factor's significance is low enough. \\
As it will be discussed in the application part, factors hypthetically may have \textbf{common} variances owing to not only from estimated pollution sources, but also from meterological and socioeconomic factors, too.

\centering\section{A Case Study for Applying PCA, ICA, FA to Identify Nitrogen Pollution Sources: Sacramento River}
\justifying

After describing overall characteristics of source apportionment of nitrogen pollution and several statistical methods used for source separation, in this chapter I will apply these algorithms to a real dataset collected by USGS from Sacramento River's site SACRAMENTO R A FREEPORT CA  coded with USGS-11447650 (USGS, 2022). This site is chosen as it has relatively high number of historical data in terms of different variables, and for future works there are ancillary data regarding point source pollution, pesticide use nearby, land use, additional water quality analyses campaigns, atmospheric deposition data.\\
For now, only one location is chosen to practically eliminate spatial autocorrelation impact. 32 parameters are filtered having more than 400 data points between year 1958 and 2022, collected from surface water (WS code). The data obtained via "dataRetrieval" package in R (R Core Team, 2022) further filtered with "not having NA for 00618" (nitrate related parameter) and later removing all rows with NA values. This resulted in 124 rows and 30 variables spanning from October 22,1985 to September 9, 2000. One last and significant step before the analyses is temporal preprocessing. The aim is getting rid of irregularly spaced data problem and removing the impact of temporal correlation. Annual means are taken to mask the impacts of monthly and seasonal dynamics. The variables with "NA" values after annual mean operation were dropped, resulted in remaining 17 variables. Deleting the variables that are directly related to each other (such as "Nitrate nitrite" in the presence of Nitrate and Nitrite) the list further trimmed down to 11 variables. Autocorrelation function for the variables in the temporal dimension did not point out major relations. Hence, this dataframe will be entered to PCA,ICA,FA methods. \\
In summary, the sources of changes in yearly differences on nitrogen water quality variables are sought after. The reason of the inclusion of non-nitrogen species and measurements are that they can assist interpretation.

\centering\subsection{PCA}
\justifying
To the difference data.frame with 38 rows and 11 variables, PCA was applied with prcomp function of R. The input data is centered, as well as scaled since there are many different scales within the variable set. Principal Components (PCs) with standard deviations higher than 1 were retained, resulted in 4 PCs.

\begin{center}
\centering\captionof{table}{PC loadings on selected variables \textit{(Un stands for unfiltered samples including solid fraction in the analysis, capital Ns at the and and Po4 means values given in "as nitrogen" and "as phosphate")}}
\begin{tabular}{ c c c c c}
\hline

Variables &PC1     &    PC2  &       PC3  &      PC4\\
Water Temperature          & -0.1755410& -0.38984238& -0.01303532 & 0.2804046\\
Dissolved Oxygen & 0.1187128 & 0.27379533 & 0.31632429& -0.4363757\\
pHUn            & -0.4619997& -0.26795873 & 0.11140634& -0.2611894\\
CarbonDioxide    & 0.2833118 & 0.38008688& -0.14124896 & 0.4847708\\
Organic N            & -0.2860378 & 0.02734127& -0.53273642& -0.3712082\\
Ammonia         & -0.3408886& -0.09170747 & 0.02860327 & 0.3678343\\
Nitrite         & -0.1451199 & 0.13136421 & 0.30351669& -0.2083277\\
Nitrate N        & -0.1543979 & 0.59988497& -0.06531265& -0.1205199\\
Phosphorus Po4   & -0.4482934 & 0.14167495 & 0.22421733 & 0.1691096\\
PhosphorusUn    & -0.2950863 & 0.23006018& -0.56526880 & 0.0667281\\
Chloride        & -0.3611361 & 0.31227442 & 0.34145247 & 0.2522957\\
\textbf{Total stdev} &\textbf{1.7091970}  &\textbf{1.4538088} &\textbf{1.2109395} &\textbf{1.0650152}\\
\hline
\label{Table. 2} 
\end{tabular}
\end{center}

\justifying
Reiterating briefly, PCA aims to find out if there is any interesting patterns in the in the loadings of the variables (annual average concentrations) under different principal components. The four PCs retained according to the Kaiser criterion together explains 64 percent of the variation in this dataset, which looks relatively low but there is no expected spatial and temporal correlation processed dataset. And considering the last row of Table. 2, already low standard deviations levels off after PC3, indicating that it would be an elbow in a scree plot.
When the results are examined while keeping aforementioned situation in mind, PC1 has, in decreasing explanatory power had high loadings for Phosphorus as phosphate, unfiltered pH, chloride and ammonia \textit{all negatively}. PC2 has distinct positive loading for Nitrate N followed by comparatively insignificant loadings. PC3 has relatively high loading of nitrite and negative loading in organic N and unfiltered phosphorus. From this much of information, it may be inferred that PC1 might reflect the variation in domestic wastewater quality or entry to the system and/or phytoplankton filtering of the water , while PC2 is a process introducing nitrate, which might be agricultural fertilizer runoff, and PC3 might be a phytoplankton existence as well with decrease in total phosphorus and increase in dissolved oxygen

\centering\subsection{ICA}

\justifying
For conducting ICA, "fastICA" package in R was used with changing number of sources 2 to 5, among them 3 did not converge to a result. The parameters used were: maximum iteration of 200 and tolerance of 0.0001, no standardization, method "C", simultaneous extraction (alg.typ = "parallel"), and logcosh functional form. 
The following Figure 2 shows the independent components together. The ICA algorithm illustrated a way to generate the variances in this study's dataset with 3 independent components. Curiously enough all these components have spikes and dips in different places. Let's first check the nitrate and ammonia annual average values and find out if there is any change in phase.

\begin{figure}
\includegraphics[scale=0.5]{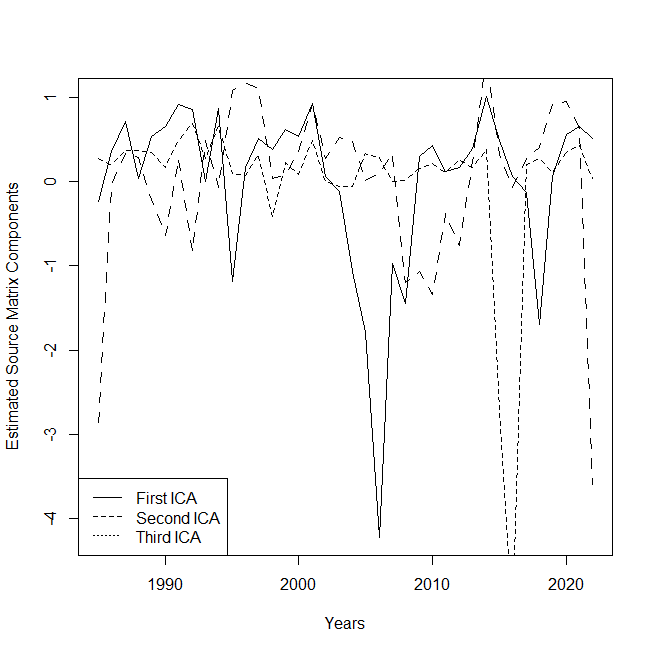}
\centering\caption{The Independent Components uncovered by ICA with 3 components}
\label{Fig. 2} 
\end{figure}

\begin{figure}
\includegraphics[scale=0.5]{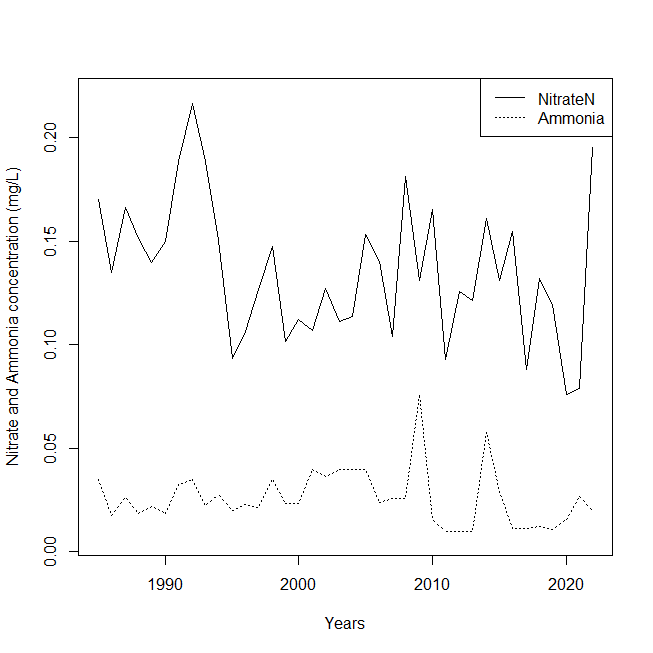}
\centering\caption{Annual average time series of nitrate and ammonia}
\label{Fig. 3} 
\end{figure}
\justifying

Direct estimation of nitrate, ammonia or their year-to-year differences linearly with these components fails. Another approach is trying to go from known contaminators to these components. Nitrogen atmospheric deposition in National Trend Network CA88 station has similar iin-phase trends to the second Independent Component. Schlegel and Domagalski's peer reviewed study (2015) has also total N fertilizer sold in Sacramento Basin between 1987-2010, which again has some spikes reminiscent of Second ICA here. Nevertheless, there are other even harder to quantify nitrogen sources, such as soil decomposition, urban runoff; and we have quite sharp uncompensateed drops in ICs for many years, which require further data/analysis, and interpretation to infer something concrete. 

\centering\subsection{FA}
\justifying
"factanal" function of R is used to conduct exploratory factor analysis on this difference data. Factor analysis was done with 1,2,3, 4 and 5 factors first, after getting the result from PCA with no rotation and maximum likelihood estimation as the method to find factors.
The "p-value" for the resulting FAs follow as in order from FA for 1 factor to FA for 5 factor: 1.17e-06, 0.0321, and 0.113.
Although 2-factor analysis was chosen to further examine the outputs, the element sof the residual matrix were too distant to being zero so the entire FA undertaking was deemed unsuccessful \textit{to explain the underlying factors impacting the changes in annual average values of the selected variables.}

\centering\section{Concluding Remarks}
\justifying

In this paper, it was sought to approach to the source apportionment of river nitrogen pollution from many perspectives of PCA and similar analyses. As suggested by several authors when PCA does not seem to be enough, ICA looks promising to solve this problem as well. The next steps are the inclusion of spatial and temporal dimensions, time-series analysis and the so-called temporal-PCA method: Singular Spectrum Analysis. This and following studies aim to include more rigorous theretical basis on the use of these dimension reduction and source separation technics for nitrogen river pollution source apportionment task.

\section{References}

\justifying

B. Everitt, T. Hothorn. An Introduction to Applied Multivariate Analysis with R. \textit{Springer}. (2011)\\
B. Schlegel, J.L. Domagalski. \textit{San Francisco Estaury and Watershed Science}, \textbf{13} (4). (2015)\\
D. Ussiri, R. Lal. Global Nitrogen Cycle, \textit{chapter in "Soil emission of Nitrous Oxide and Its Mitigation} 29-62. (2012)\\
I. T. Joliffe. Principal Component Analysis. \textit{Springer}. 2nd Edition. (2002).\\
J. N. Galloway, \textit{Science in China Series C: Life Sciences}, \textbf{48}, 669-678. (2005)\\
J. Shlens. A Tutorial on Principal Component Analysis. (2014). arXiv:1404.1100 \\
J. Shlens. A Tutorial on Independent Component Analysis. (2014). arXiv:1404.2986 \\
K. Adachi. Matrix Based Introduction to Multivariate Data Analysis. \textit{Springer}. (2016)\\
R Core Team. R:  Language and Environment for Statistical Computing. \textit{R Foundation for Statistical Computing}. (2022).\\
USGS. USGS 11447650 Sacramento R A Freeport CA. url:https://waterdata.usgs.gov/ (2022).\\

\end{document}